\documentclass[11pt]{article}
\usepackage[utf8]{inputenc}
\usepackage{listings}
\usepackage{color}
\usepackage{graphicx}
\usepackage{longtable}
\usepackage{float}
\usepackage{amsmath}
\usepackage{amssymb}
\usepackage{authblk}
\usepackage{outlines}
\usepackage{subcaption}
\usepackage{todonotes}

\title{}
\date{December 9, 2024}
\begin{document}

\title{Timing consistency of T cell receptor activation in a stochastic model combining kinetic segregation and proofreading}
\author{Thorsten Pr\"ustel} 
\author{Martin Meier-Schellersheim} 
\affil{Computational Systems Biology Section\\Laboratory of Immune System Biology\\National Institute of Allergy and Infectious Diseases\\National Institutes of Health, Bethesda, Maryland 20892, USA}
\maketitle
\let\oldthefootnote\thefootnote 
\renewcommand{\thefootnote}{\fnsymbol{footnote}} 
\footnotetext[1]{Email: prustelt@niaid.nih.gov} 
\let\thefootnote\oldthefootnote

\begin{abstract}
T cell receptor signaling must operate reliably under tight time constraints. While assuming quite different mechanisms, two prominent models of T cell receptor activation, kinetic segregation and kinetic proofreading, both introduce a distinct time scale. However, a clear understanding of whether and how those characteristic times give rise to a consistent timing of T cell receptor activation in the presence of stochastic fluctuations has been lacking so far. Here, using a simulation approach capable of modeling molecular interactions between adjacent cell membranes, we explore a stochastic model that combines elements of kinetic segregation and proofreading. Our simulations suggest that the two mechanisms interoperate, thereby rendering the corresponding stochastic times biologically functional. Receptor activation relies on rare molecular events that are not well characterized by the mean of the underlying probability density function. Yet, a consistent timing of receptor activation can be ensured by a modest number of proofreading steps.
\end{abstract}

\section{Introduction}
\label{introduction} 
T cells use the T cell antigen receptor (TCR) to scan the surface of antigen-presenting cells (APCs) and other cells to detect antigens in the form of peptide fragments bound to major histocompatibility complex (MHC) class I or class II proteins. Engagement with peptide:MHC (pMHC) complexes initiates TCR signaling, a prerequisite for T cell activation and mounting an adaptive immune response \cite{chakraborty2014insights}. 

A substantial body of experimental work has revealed a unique combination of TCR activation hallmarks. T cells' capacity for ligand discrimination is characterized by an exquisite specificity, extraordinary sensitivity and speed \cite{feinerman2008quantitative, malissen2015early}. Foreign (or non-self) pMHCs are presented next to an orders of magnitude larger number of endogenous (or self) pMHCs. The T cell's task of selective pMHC discrimination is further compounded by the typically small differences in affinity and kinetics of binding to foreign and self pMHC. Yet, studies suggest that a few and even a single agonist pMHC are sufficient to initiate T cell activation \cite{irvine2002direct, huang2013single}. Furthermore, to enable efficient scanning of APCs, one would expect that the decision-making of the TCR signaling apparatus has to operate under tight time constraints \cite{brodovitch2013t, cai2017visualizing} and, indeed, it has been shown that phosphorylation of TCR-proximal signaling components and onset of calcium influx occur after only a few seconds \cite{houtman2005early, huse2007spatial}.

Despite the amassed experimental data and no shortage of theoretical models \cite{van2011mechanisms, lever2014phenotypic, chakraborty2014insights, malissen2015early, courtney2018tcr}, so far no conceptual basis has emerged that fully meets the challenge to reconcile the aforementioned distinct features and that provides a quantitative detailed picture of the molecular mechanisms that underlie TCR signal initiation. However, several models have been proposed that can at least account for some of the observed TCR activation behaviors.

Kinetic proofreading (KP) has originally been conceived to explain the observed low error rates of DNA replication that are not compatible with the size of fluctuations in thermodynamic equilibrium. The basic idea is to modify the equilibrium binding scheme by adding $N$ irreversible kinetic steps. Then, the error rate of the resulting non-equilibrium process is the $N$-th power of the equilibrium error rate and in the limit of large $N$ the error rate becomes arbitrarily small.
KP has quickly found applications beyond its initial context, for instance in signal transduction and, in particular, in TCR ligand discrimination \cite{mckeithan1995kinetic} and is nowadays considered to be a general design principle of many biochemical networks \cite{alon2019introduction}. However, while it can in principle give rise to arbitrarily high specificity, in its simplest form it is at odds with the two other hallmarks, speed and sensitivity.
The requirement of a large number of proofreading steps renders it a slow process and each step attenuates the signal strength. 

These issues have been known to plague KP models for a long time. In fact, already McKeithan's seminal paper \cite{mckeithan1995kinetic} raises the issue of signal strength attenuation. More recently, theoretical evidence suggests that KP fails even at its \textit{raison d'etre}, because it leads to a degraded ligand resolution performance, when the impact of molecular noise is accounted for \cite{kirby2023proofreading}. The apparent improvement of ligand discrimination is only realized in deterministic versions of KP that describe averaged quantities and ignore stochastic fluctuations. 
These results are consistent with an analysis of the timing characteristics of activation in KP  models \cite{bel2009simplicity}. It was shown that different versions of KP either exhibit a clock-like, deterministic or fully stochastic (exponential) behavior as the number of proofreading steps increases. However, the KP scheme that serves as a blueprint for many models of TCR signaling is more likely than not to behave stochastically \cite{bel2009simplicity}.

These findings raise the question to what extent KP can still be considered as a viable model of the TCR's high selectivity. Adding to this conundrum, recent direct experimental tests of KP suggest that the effective number of KP steps lies between 2-3 \cite{tischer2019light}. This is a surprisingly low number from the perspective of KP models that seek to explain high specificity, but it is compatible with the aforementioned theoretical work that shows that the degrading influence of fluctuations can be avoided with a small number of proofreading steps. 

Alternatives models of TCR signaling initiation have been suggested that build on the special morphological features of the contact sites formed by T cells and APCs during antigen scanning and the ensuing constraints on the local biochemistry. According to the Kinetic Segregation (KS) model \cite{chang2016initiation, davis2006kinetic} the large ectodomain of CD45 and other phosphatases leads to their passive exclusion from the sites of TCR engagement. Thus, locally the spatial distribution of membrane complexes is altered and regions with diminished phosphatase activity emerge that facilitate the phosphorylation of TCR-proximal signaling components. In a similar vein, a more recently studied model suggests that TCR activation is essentially regulated by the TCRs dwell time in phosphatase depleted cell-cell close contact sites \cite{fernandes2019cell}. The dwell time characteristics is modulated by several factors, such as TCR-pMHC interactions, the membrane molecules' diffusive motion and the size of the contact region. A further postulate is that any TCR remaining within the contact region for longer than an ad-hoc postulated critical time gets activated.

On the face of it, the KP and KS model appear to be quite different. While KP is a purely biochemical process, the KS scheme relies heavily on spatial aspects. Yet, both mechanisms share a common key characteristic: they introduce a random time \cite{chakraborty2014insights}. Random times, such as waiting times and first-passage times, feature prominently in the theory of stochastic processes \cite{van2011stochastic} and have found applications in, among many diverse fields, cell biology \cite{iyer2016first, redner2023first}. Waiting times record the duration that it takes for an event to occur. When, in particular, the first occurrence of the event is of interest, waiting times are also referred to as first-passage times \cite{van2011stochastic, iyer2016first, redner2023first, huang2021relating}. In the KP scheme, the random time is the time that elapses before the sequence of proofreading steps is completed. In the KS model, the random time is given by the TCRs dwell time, also referred to as first-exit, or exit time hereafter. We note that for spatially homogeneous systems in thermodynamic equilibrium both random times cease to exist. Hence, the rationale for driving the system far from equilibrium (KP) and reorganizing the spatial distribution of membrane proteins (KS) can also be understood from a random time point of view. 

The experimental evidence that both the KP and KS scheme play a role in the initiation of TCR signaling raises questions about the relationship between the two timescales set by the respective random times. Do they 'conspire' to establish and orchestrate the functional timescale of TCR activation? If so, how can swift activation be achieved, given the intrinsic slowness of the KP? Also, how is timing consistency ensured in the presence of stochastic fluctuations that, as mentioned, diminish KP's capacity for ligand resolution? 

To create computational models that could help address these questions, it follows from the discussion above that simulation methods are required that incorporate stochastic approaches to biochemistry into computational representations of cellular morphology. In particular, models of TCR-pMHC interactions between two adjacent membranes should be capable of accounting for the morphological and biochemical features that distinguish cell-cell contact sites from other regions of the cell membrane.
Such a simulation approach has been introduced recently \cite{prustel2023grid} and provides the basis of our computational exploration of the interplay between the contact site's morphological features and the stochasticity of the TCR's interactions and their role in shaping TCR activation.

We note that the timing characteristics of cellular signaling processes, such as Ras activation by SOS and response times of cell-to-cell communication networks, have been studied in terms of first-passage times before \cite{thurley2018modeling, huang2019molecular, huang2021relating}. Here, we study a stochastic single-particle based spatially resolved model of TCR activation that accounts for the impact of both cell morphology and fluctuations.

The remainder of this manuscript is structured as follows. In Sec.~\ref{Model} we describe a stochastic model of TCR activation that combines elements of KS and KP. Sec.~\ref{sec:waiting_times} describes the mathematical quantities and relations that account for the activation time distribution of a single TCR and the corresponding activation probabilities. The simulation setup is detailed in Sec.~\ref{sec:sim_setup}. In Sec.~\ref{sec:results} we present the simulation results and conclude with Sect.~\ref{sec:conclusion}.

\section{Stochastic model of T cell receptor activation}\label{Model}
The model we explored combines elements of both KS and KP. The former asserts that due to their large ectodomain, phosphatases, such as CD45, are excluded from contact sites formed by T cell and APC, thereby enabling phosphorylation of those TCRs that reside at these specialized sites of cell-cell contact.  A recent extension proposes that any TCR whose dwell time in the contact region exceeds an ad-hoc postulated reference time becomes activated \cite{fernandes2019cell}. Apart from TCR activation, KS does not specify any details on what should happen within the dwell time.

KP, by contrast, refrains from considering potential morphological influences on the local biochemistry and, instead, asserts that a ligated TCR must complete a sequence of biochemical transformation before it dissociates to reach the fully activated state. 

Hence, both KS and KP critically rely on the existence of a stochastic ‘clock’, the dwell and completion time, respectively. Here, we propose that the proofreading’s completion time can serve as the reference clock in the KS scheme, thus establishing together with the dwell time the TCR activation time scale. In this way, the proofreading mechanism also fills the aforementioned gap of the KS model and describes in more detail what happens within the dwell time.
\subsection{Random times and activation probabilities}\label{sec:waiting_times}
The model accounts for the timing of the TCR activation by introducing three random times, $\tau_{\text{act}}, \tau_{\text{cmpl}}$ and $\tau_{\text{exit}}$.
These stochastic quantities represent the time it takes a TCR to arrive in the activated state, to complete the sequence of proofreading steps and to exit the contact site, respectively. Hereafter, we use the terms dwell time and exit time interchangeably. The relationship between these three random times is expressed in terms of the corresponding probability measures $\Pr$ as follows \cite{huang2016phosphotyrosine}
\begin{equation}
\label{prob_3_times}
\Pr(\tau_{\text{act}} \leq t) = \Pr(\tau_{\text{cmpl}} \leq\tau_{\text{exit}}) \Pr(\tau_{\text{cmpl}} \leq t)
\end{equation}
Eq.~(\ref{prob_3_times}) expresses the essence of the model, namely that to get activated, a TCR has to complete the proofreading steps before it leaves the contact site.
Recalling that the cumulative probability function (CDF) of a random time $\tau$ is given by
$P_{\tau}(t) = \Pr(\tau \leq t)$
and related to the corresponding probability density function (PDF) via
$p_{\tau}(t) = \frac{d P_{\tau}(t)}{dt}$,
we arrive from Eq.~(\ref{prob_3_times}) at
\begin{equation}
\label{p_act}
p_{\tau_{\text{act}}}(t) = \Pr(t < \tau_{\text{exit}}) p_{\tau_{\text{cmpl}}}(t),
\end{equation}
where $\Pr(t < \tau_{\text{exit}}) = 1 - \Pr(\tau_{\text{exit}} \leq t)$.

The activation probability $P_{\text{act}}$ of a single TCR is given by
\begin{equation}
\label{P_act_single}
P_{\text{act}} = \int^{\infty}_{0} p_{\tau_{\text{act}}}(t)\,dt.
\end{equation}

Assuming that all $N$ proofreading steps are irreversible and characterized by the same rate constant $k_{p}$, the completion time PDF $p_{\tau_{\text{cmpl}}}(t)$ assumes the form of a gamma density function 
\begin{equation}
\label{gamma_density}
p_{\tau_{\text{cmpl}}}(t) = \frac{k_{p}^{N}t^{N-1}}{(N-1)!}\exp(-k_{p}t).
\end{equation}
By contrast, no analytical expression for $\Pr(t < \tau_{\text{exit}})$ is known. This probability depends on several features, such as the geometry of the contact site, the TCR's and pMHC's diffusion constant, the kinetic parameters of the TCR-pMHC interactions and the pMHCs membrane density. Clearly, this probability can be obtained by simulation only.

Knowledge of the activation probability $P_{\text{act}}$ of a single TCR (Eq.~(\ref{P_act_single})) permits to calculate the probability of exactly $k$ activated TCRs. Considering every TCR that enters the region of close contact as an independent 'trial', this probability is given by the binomial distribution 
\begin{equation}
\label{P_k}
P(k; n, P_{\text{act}}) = \binom{n}{k} P_{\text{act}}^{k}(1-P_{\text{act}})^{n-k},
\end{equation}
where $n$ is the total number of TCRs that entered the contact site. 

Finally, we assume that T cell activation requires a certain threshold number $K$ of activated TCRs. Using Eq.~(\ref{P_k}), the probability that at least $K$ TCRs get activated is given by
\begin{equation}
\label{P_at_least_K}
P(k \geq K; n, P_{\text{act}}) = 1 - \sum^{K-1}_{i=0}P(i; n, P_{\text{act}}).
\end{equation} 

In the limit of small $P_{\text{act}}$ and large $n$, the binomial distribution is well approximated by the Poisson distribution $\lambda^{k}/k!e^{-\lambda}$, where $\lambda=n P_{\text{act}}$ is the mean number of activated TCRs. Making the ansatz that $\lambda$ is proportional to the pMHC surface density, $\lambda = \alpha [\text{pMHC}]$, and using Eq.~(\ref{P_at_least_K}), one arrives at
\begin{equation}
\label{fit_func}
P(k \geq K; [\text{pMHC}]) = 1 - \exp(-\alpha [pMHC])\sum^{K-1}_{i=0}\frac{(\alpha [\text{pMHC}])^{i}}{i!}.
\end{equation} 
\subsection{Simulation setup}\label{sec:sim_setup}
To numerically construct the probability densities $p_{\tau_{\text{exit}}}(t), p_{\tau_{\text{act}}}(t)$ of the dwell and activation times, we applied a single-particle based stochastic simulation approach that has been introduced recently \cite{prustel2023grid}. It can track the stochastic motion and interaction of individual molecules, including those that are located on separate, adjacent membranes. Its capability to trace individual TCRs permits to obtain realizations of the corresponding random times $\tau_{\text{exit}}, \tau_{\text{act}}$ as a direct readout. The simulation method employs a smooth and grid-free, so-called blob- or metaball-based computational representation of cellular morphologies that is compatible with particle-based stochastic approaches. Blobby modeling offers simple and efficient means of generating realistic geometries and implementing constraints, such as the ones imposed by the cell-cell contact site’s morphology on the diffusion of membrane complexes.

As in Ref.~\cite{prustel2023grid}, we simulated a filopodia-like protrusion on a T cell that forms an interface with an APC. The pMHC expression levels and two-dimensional (2D) affinities for binding to TCR were varied ($[\text{pMHC}]=1,\ldots, 300\mu m^{-2}$, APC area $\sim 360\mu m^{2}$, $\kappa_{\text{off}}=1,\ldots, 50s^{-1}$, $\kappa_{\text{on}}=0.1\mu m^{2}s^{-1}$). Here, the 2D on-and off-rates refer to the microscopic rates of the Smoluchowski-Collins-Kimball approach to diffusion-influenced reactions \cite{collins1949diffusion}.
We defined the cell-cell contact site as the the protrusion’s tip's region from which membrane complexes with a steric height of $h_{steric}=25 nm$ are excluded due to their size. The resulting contact site radius is $r_{c}=214nm = \sqrt{A_{c}/\pi}$, where $A_{c}$ refers to the contact site's area. The other used parameters are the diffusion coefficients $D_{\text{TCR}} = D_{\text{pMHC}} = 0.05 \mu m^{2}s^{-1}$, TCR membrane density $[TCR] = 100\mu m^{-2}$, T cell area $\sim 487 \mu m^{2}$ and cell-cell contact duration $t_{sim} = 30s$. At the start of each simulation run, both TCR and pMHC homogeneously covered the complete respective cells surface.

 When a TCR entered the contact site, the entry time was recorded and the proofreading sequence started. When a TCR left the contact region, its dwell-time was recorded, the associated proofreading sequence stopped and the current proofreading state returned to the initial proofreading state. Finally, when a TCR completed the proofreading sequence before leaving the region of contact, the activation time was recorded. 
 \begin{figure}
 \begin{subfigure}[t]{0.5\textwidth}
\includegraphics[scale=0.3]{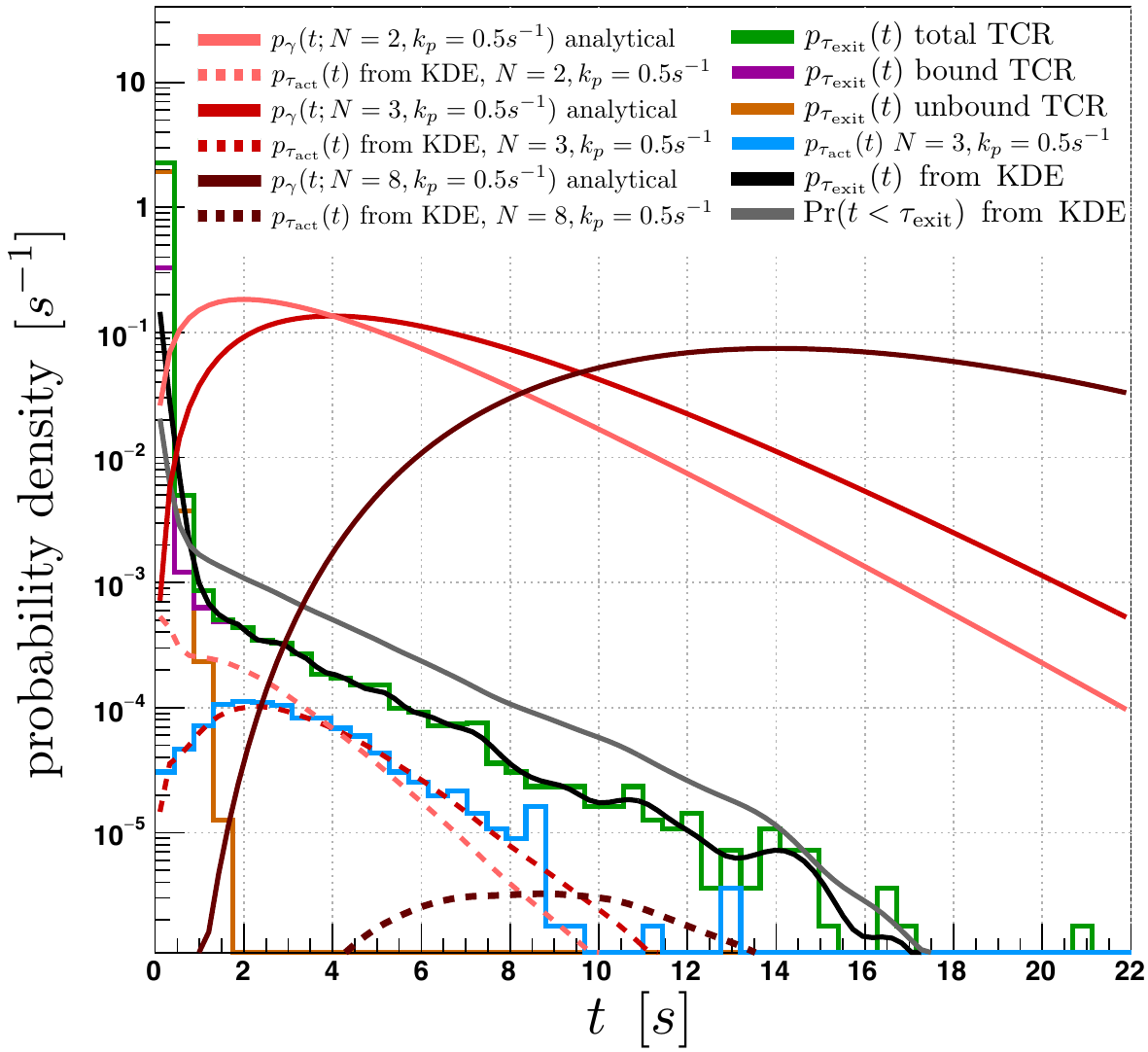}
\caption{
Relationship between completion, dwell and activation time probability density functions.
$[\text{pMHC}] = 3\mu m^{-2}, k_{\text{off}}= 1 s^{-1}$.
}
\label{sub_fig:act_time_pdf}
\end{subfigure}
~
\begin{subfigure}[t]{0.5\textwidth}
\includegraphics[scale=0.3]{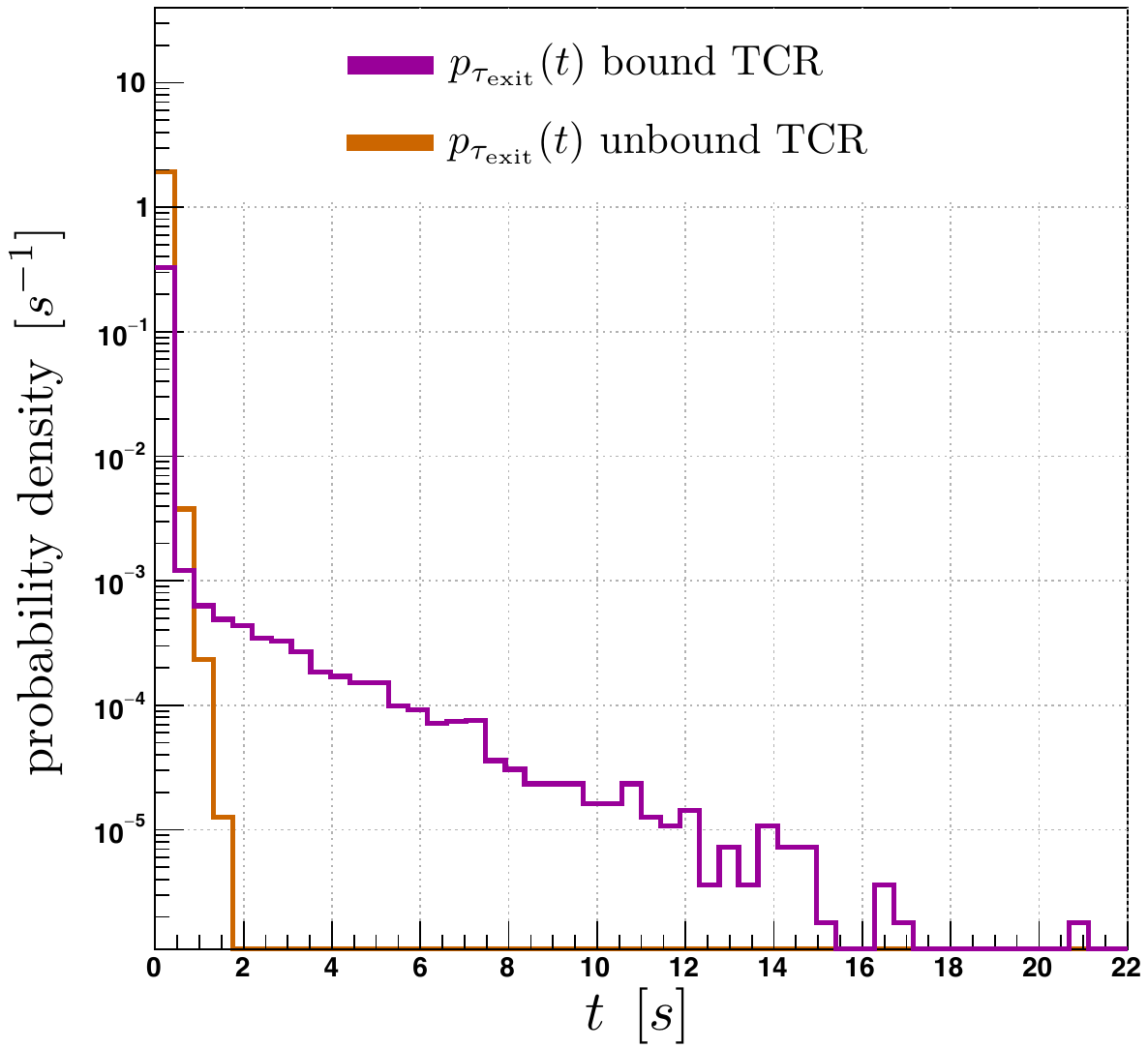}
\caption{
The dwell-time PDF depends
on whether a TCR associated with a pMHC. $[\text{pMHC}] = 3\mu m^{-2}, k_{\text{off}}= 1 s^{-1}$.
}
\label{sub_fig:unbound_vs_bound_pdf}
\end{subfigure}
\caption{
TCR dwell and activation time probability densities.
}\label{fig:random_time_pdf}
\end{figure}
\section{Results}\label{sec:results}
The semi-log plot of the TCR dwell time probability density exhibits a long right tail, suggesting at least two different timescales corresponding to two TCR subpopulations (Fig.~\ref{sub_fig:act_time_pdf}, green solid line). Fig.~\ref{sub_fig:unbound_vs_bound_pdf} shows that the long tail arises virtually exclusively from the TCRs that at some point bind to pMHC at least once and subsequently experience diffusional trapping \cite{chen2021trapping}. Therefore, the molecular events leading to TCR activation are rare events associated with long dwell times that substantially deviate from the PDF's mean. Furthermore, the squared coefficient of variation $\text{CV}^{2}$ of the dwell time PDF exceeds unity by far $\text{CV}^{2}\gg 1$. $\text{CV}^{2}$ is also known as randomness parameter. It assumes unity for an exponential distribution, while values smaller than one indicate a more deterministic process \cite{huang2021relating}. Hence, the question arises how timing consistency can be ensured under these circumstances.

Because the TCRs dwell times give rise to at least two different timescales, a standard molecular clock is required that can distinguish them to render them biologically functional. As suggested in Sec.~\ref{Model}, KP and its associated completion time appear as a natural candidate for such a temporal measure. 
Fig.~\ref{sub_fig:act_time_pdf} shows that the KP's PDF (red solid line) substantially overlaps with dwell time distribution of the TCRs that did bind to pMHC at least once.The corresponding simulated activation time PDF is indicated by the blue solid line.
The precise interplay of the KS's and KP's specific random times and TCR activation times described by Eq.~(\ref{p_act}) permits to check the simulation result. We can numerically construct $p_{\tau_{\text{act}}}(t)$ in an alternative manner that relies on $p_{\tau_{\text{exit}}}(t)$ and the analytical expression of $p_{\tau_{\text{cmpl}}}(t)$. First, we performed a Kernel Density Estimation (KDE) to obtain an estimate for $p_{\tau_{\text{cmpl}}}(t)$ and, in turn, of $\text{Pr}(\tau_{\text{exit}} < t)$ (Fig.~\ref{sub_fig:act_time_pdf}, black and grey lines, respectively). The KDE representation facilitates multiplication with the exact expression of $p_{\tau_{\text{cmpl}}}(t)$ (Fig.~\ref{sub_fig:act_time_pdf}, red line) to finally arrive at an estimate for $p_{\tau_{\text{act}}}(t)$. We find good agreement with the direct readout from the simulation (Fig.~\ref{sub_fig:act_time_pdf}, red dashed line vs blue solid line).

The KDE construction opens the possibility to reuse the dwell time PDF to efficiently explore how the distribution of TCR activation times changes with varying number $N$ of proofreading steps. Fig.~\ref{sub_fig:act_time_pdf} shows the resulting activation time PDFs that shift to later times with increasing $N$ (dashed lines). The results suggest that KP is necessary to establish a realistic activation timescale and to filter out spurious short dwell events. Very short or no proofreading ($N\leq 2$, where $N=1$ corresponds to an exponentially distributed completion time, that is, no proofreading) result in too fast activations, because spurious short dwell time events are not filtered out. Another related consequence is that a substantial fraction of activated TCRs belongs to the subpopulation that did not bind to pMHC at all.

By contrast, longer proofreading reduces the overlap of the completion time pdf with the dwell time histogram, thereby ensuring that virtually only long dwell time events associated with TCRs that did bind to pMHC contribute to the activation time pdf. However, this also means that as $N$ increases, TCR activations and activation probabilities (Eq.~(\ref{P_act_single})) become increasingly slow and small, respectively.

The mean activation time is different from both the mean of the dwell and the mean of the completion time pdf for a given $N$ and lies between these two values. The reason is that activations arise from the subset of rare long TCR dwell time events that coincide with fast completions. The same mechanism underlies the timing consistency of the activation time PDF and explains why it does not share the features of the dwell-time density, such as $CV^{2}\gg 1$ and a long tail. 

Finally, turning from activation of single TCRs to T cell activation, we defined the probability thereof as the probability of activation of at least 10 TCRs within a time window of 30s. Fig.~\ref{fig:P_act_vs_pMHC}, obtained by fitting Eq.~\ref{fit_func} to the simulation results, shows the dependence of T cell activation on pMHC density for varying off-rates. While the combination of KS and KP results in a good discrimination between agonist and non-agonist ligands (corresponding to 2D off-rates $\sim 10s^{-1}$ \cite{liu2014accumulation}), the achieved separation does not amount to the exquisite specificity reported in the literature \cite{feinerman2008quantitative} and does not prevent activation by high densities of low-affinity pMHC. 


\section{Discussion}\label{sec:conclusion}
We explored a stochastic spatially-resolved computational model of TCR activation that combines elements of KS and KP. A prerequisite for our study was the capability of the applied simulation approach to account for the morphological features of the TCR-APC contacts and to track individual TCRs \cite{prustel2023grid}, thereby permitting to record dwell and activation times. Our results suggest that KP provides a molecular clock to separate the two TCR subpopulations that give rise to the dwell time PDF. 

The observed dependency on $N$ shows that KP is necessary to establish a realistic distribution of activation times and to filter out spurious dwell time events. It also suggests that a modest number of proofreading steps is sufficient to achieve these effects, while many proofreading steps lead to less realistic activation times and probabilities. This is consistent with previous experimental estimates of $\sim$3 proofreading steps for TCR activation \cite{tischer2019light}.

\begin{figure}
\includegraphics[scale=0.5]{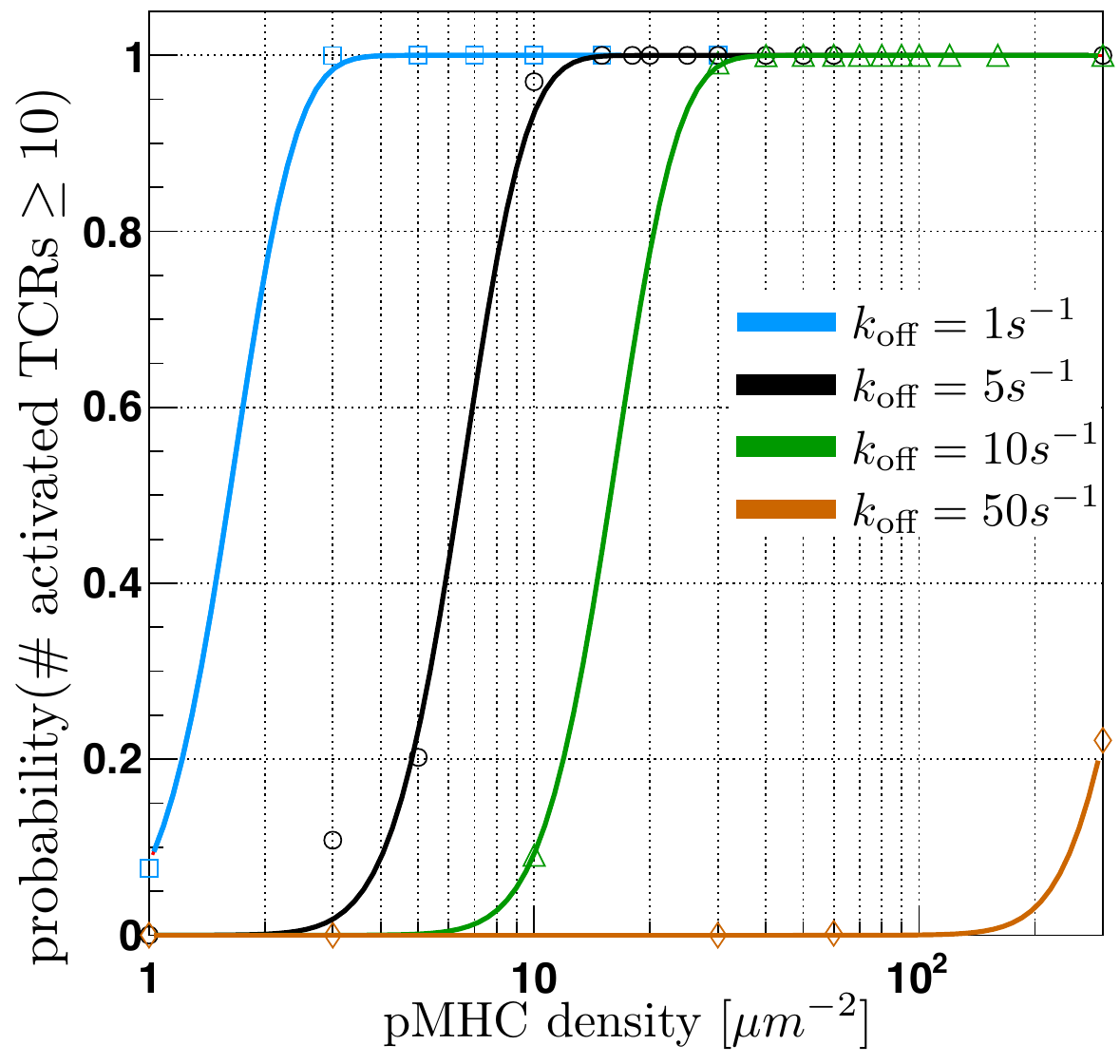}
\caption{
Activation probability of at least 10 TCR vs the pMHC density for various 2D off-rates.
}
\label{fig:P_act_vs_pMHC}
\end{figure}

The achieved good but not exquisite specificity warrants an exploration of signaling mechanism that could replace the abstract KP model with more realistic models of intracellular biochemistry and increase ligand discrimination. Furthermore, we note that Fig.~\ref{fig:P_act_vs_pMHC} is based on Eq.~(\ref{fit_func}) that, in turn, assumes that every TCR that enters the contact site represents an independent trial. This assumption may not be valid, for instance, if the dwell time distribution is not strictly stationary. Finally, a better understanding of the relationship between experimentally derived kinetic rates that govern TCR-pMHC binding and their quantitative definition is required \cite{faro2017unifying}.  
Simulation approaches that can account for fluctuations, cell morphology and biochemistry will help address these questions.

\section*{Acknowledgments}
This research was supported by the Division of Intramural Research of \mbox{NIAID, NIH.} 

\bibliographystyle{plain} 

\end{document}